# Measurement of Cabibbo-suppressed Decays of the $\tau$ Lepton

M. Battle,[1] J. Ernst,[1] Y. Kwon,[1] S. Roberts,[1] E.H. Thorndike,[1] C.H. Wang,[1] J. Dominick,[2] M. Lambrecht,[2] S. Sanghera,[2] V. Shelkov,[2] T. Skwarnicki,[2] R. Stroynowski,[2] I. Volobouev,[2] G. Wei,[2] P. Zadorozhny,[2] M. Artuso,[3] M. Goldberg,[3] D. He,[3] N. Horwitz,[3] R. Kennett,[3] R. Mountain,[3] G.C. Moneti,[3] F. Muheim,[3] Y. Mukhin,[3] S. Playfer,[3] Y. Rozen,[3] S. Stone,[3] M. Thulasidas,[3] G. Vasseur,[3] G. Zhu,[3] J. Bartelt,[4] S.E. Csorna,[4] Z. Egyed,[4] V. Jain,[4] K. Kinoshita,[5] K.W. Edwards,[6] M. Ogg,[6] D.I. Britton,[7] E.R.F. Hyatt,[7] D.B. MacFarlane,[7] P.M. Patel,[7] D.S. Akerib,[8] B. Barish,[8] M. Chadha,[8] S. Chan,[8] D.F. Cowen,[8] G. Eigen,[8] J.S. Miller,[8] C. O'Grady,[8] J. Urheim,[8] A.J. Weinstein,[8] D. Acosta,[9] M. Athanas,[9] G. Masek,[9] H.P. Paar,[9] M. Sivertz,[9] J. Gronberg,[10] R. Kutschke,[10] S. Menary,[10] R.J. Morrison,[10] S. Nakanishi,[10] H.N. Nelson,[10] T.K. Nelson,[10] C. Qiao,[10] J.D. Richman,[10] A. Ryd,[10] H. Tajima,[10] D. Sperka,[10] M.S. Witherell,[10] M. Procario,[11] R. Balest,[12] K. Cho,[12] M. Daoudi,[12] W.T. Ford,[12] D.R. Johnson,[12] K. Lingel,[12] M. Lohner,[12] P. Rankin,[12] J.G. Smith,[12] J.P. Alexander,[13] C. Bebek,[13] K. Berkelman,[13] K. Bloom,[13] T.E. Browder[†,13] D.G. Cassel,[13] H.A. Cho,[13] D.M. Coffman,[13] P.S. Drell,[13] R. Ehrlich,[13] P. Gaidarev,[13] R.S. Galik,[13] M. Garcia-Sciveres,[13] B. Geiser,[13] B. Gittelman,[13] S.W. Gray,[13] D.L. Hartill,[13] B.K. Heltsley,[13] C.D. Jones,[13] S.L. Jones,[13] J. Kandaswamy,[13] N. Katayama,[13] P.C. Kim,[13] D.L. Kreinick,[13] G.S. Ludwig,[13] J. Masui,[13] J. Mevissen,[13] N.B. Mistry,[13] C.R. Ng,[13] E. Nordberg,[13] J.R. Patterson,[13] D. Peterson,[13] D. Riley,[13] S. Salman,[13] M. Sapper,[13] F. Würthwein,[13] P. Avery,[14] A. Freyberger,[14] J. Rodriguez,[14] R. Stephens,[14] S. Yang,[14] J. Yelton,[14] D. Cinabro,[15] S. Henderson,[15] T. Liu,[15] M. Saulnier,[15] R. Wilson,[15] H. Yamamoto,[15] T. Bergfeld,[16] B.I. Eisenstein,[16] G. Gollin,[16] B. Ong,[16] M. Palmer,[16] M. Selen,[16] J. J. Thaler,[16] A.J. Sadoff,[17] R. Ammar,[18] S. Ball,[18] P. Baringer,[18] A. Bean,[18] D. Besson,[18] D. Coppage,[18] N. Copty,[18] R. Davis,[18] N. Hancock,[18] M. Kelly,[18] N. Kwak,[18] H. Lam,[18] Y. Kubota,[19] M. Lattery,[19] J.K. Nelson,[19] S. Patton,[19] D. Perticone,[19] R. Poling,[19] V. Savinov,[19] S. Schrenk,[19] R. Wang,[19] M.S. Alam,[20] I.J. Kim,[20] B. Nemati,[20] J.J. O'Neill,[20] H. Severini,[20] C.R. Sun,[20] M.M. Zoeller,[20] G. Crawford,[21] C. M. Daubenmier,[21] R. Fulton,[21] D. Fujino,[21] K.K. Gan,[21] K. Honscheid,[21] H. Kagan,[21] R. Kass,[21] J. Lee,[21] R. Malchow,[21] Y. Skovpen[⋆,21] M. Sung,[21] C. White,[21] F. Butler,[22] X. Fu,[22] G. Kalbfleisch,[22] W.R. Ross,[22] P. Skubic,[22] J. Snow,[22] P.L. Wang,[22] M. Wood,[22] D.N. Brown,[23] J.Fast ,[23] R.L. McIlwain,[23] T. Miao,[23] D.H. Miller,[23] M. Modesitt,[23] D. Payne,[23] E.I. Shibata,[23] I.P.J. Shipsey,[23] P.N. Wang,[23]

(CLEO Collaboration)

[†]Permanent address: University of Hawaii at Manoa
[⋆]Permanent address: INP, Novosibirsk, Russia

1. *University of Rochester, Rochester, New York 14627*
2. *Southern Methodist University, Dallas, Texas 75275*
3. *Syracuse University, Syracuse, New York 13244*
4. *Vanderbilt University, Nashville, Tennessee 37235*
5. *Virginia Polytechnic Institute and State University, Blacksburg, Virginia, 24061*
6. *Carleton University, Ottawa, Ontario K1S 5B6 and the Institute of Particle Physics, Canada*
7. *McGill University, Montréal, Québec H3A 2T8 and the Institute of Particle Physics, Canada*
8. *California Institute of Technology, Pasadena, California 91125*
9. *University of California, San Diego, La Jolla, California 92093*
10. *University of California, Santa Barbara, California 93106*
11. *Carnegie-Mellon University, Pittsburgh, Pennsylvania 15213*
12. *University of Colorado, Boulder, Colorado 80309-0390*
13. *Cornell University, Ithaca, New York 14853*
14. *University of Florida, Gainesville, Florida 32611*
15. *Harvard University, Cambridge, Massachusetts 02138*
16. *University of Illinois, Champaign-Urbana, Illinois, 61801*
17. *Ithaca College, Ithaca, New York 14850*
18. *University of Kansas, Lawrence, Kansas 66045*
19. *University of Minnesota, Minneapolis, Minnesota 55455*
20. *State University of New York at Albany, Albany, New York 12222*
21. *Ohio State University, Columbus, Ohio, 43210*
22. *University of Oklahoma, Norman, Oklahoma 73019*
23. *Purdue University, West Lafayette, Indiana 47907*

## ABSTRACT

Branching ratios for the dominant Cabibbo-suppressed decays of the $\tau$ lepton have been measured by CLEO II in $e^+e^-$ annihilation at CESR ($\sqrt{s} \sim 10.6$ GeV) using kaons with momenta below 0.7 GeV/c. The inclusive branching ratio into one charged kaon is $(1.60 \pm 0.12 \pm 0.19)\%$. For the exclusive decays, $B(\tau^- \to K^-\nu_\tau) = (0.66 \pm 0.07 \pm 0.09)\%$, $B(\tau^- \to K^-\pi^0\nu_\tau) = (0.51 \pm 0.10 \pm 0.07)\%$, and, based on three events, $B(\tau^- \to K^-2\pi^0\nu_\tau) < 0.3\%$ at the 90% confidence level. These represent significant improvements over previous results. $B(\tau^- \to K^-\pi^0\nu_\tau)$ is measured for the first time with exclusive $\pi^0$ reconstruction.

Measurements of Cabibbo-suppressed decays of the $\tau$ lepton provide a test of the Standard Model for the strange sector of the weak hadronic current. Of the decays with one charged kaon in the final state, only $\tau^- \to K^-\nu_\tau$ has previously been observed;[1] its branching ratio is predicted[2] to be $(0.76 \pm 0.03)\%$ based on measurements[3] of the Cabibbo-favored decay $\tau^- \to \pi^-\nu_\tau$. The decay $\tau^- \to K^-\pi^0\nu_\tau$ is expected to proceed through the $K^*(892)$ resonance and its branching ratio can be calculated[2] from the measured branching ratio[3] for $\tau^- \to \rho^-\nu_\tau$, giving $B(\tau^- \to K^-\pi^0\nu_\tau) = (0.38 \pm 0.01)\%$. The decay $\tau^- \to K^{*-}(892)\nu_\tau$ has previously been observed by reconstructing a $K_S \to \pi^+\pi^-$ accompanying a charged particle. The decay $\tau^- \to K^-\pi^0\nu_\tau$ provides an alternative measurement of the branching ratio for $\tau^- \to K^{*-}(892)\nu_\tau$ with minimal contamination from $\tau^- \to K^-K^0\nu_\tau$. There are no firm predictions for other Cabibbo-suppressed decays. Presented in this Letter are a new measurement of the branching ratio for $\tau^- \to K^-\nu_\tau$ with good precision, the first measurement of the decay $\tau^- \to K^-\pi^0\nu_\tau$ with exclusive $\pi^0$ reconstruction, and the first limit on the decay $\tau^- \to K^-2\pi^0\nu_\tau$.

The data used in this analysis have been collected with the CLEO II detector[4] at the Cornell Electron Storage Ring (CESR). CLEO II is a general purpose spectrometer with excellent charged particle and shower energy detection. The momenta and specific ionization (dE/dx) of charged particles are measured with three cylindrical drift chambers between 5 and 90 cm from the $e^+e^-$ interaction point (IP), with a total of 67 layers. These are surrounded by a scintillation time-of-flight (TOF) system and a CsI(Tl) calorimeter with 7800 crystals. These detector systems are installed inside a 1.5 T superconducting solenoidal magnet, surrounded by a proportional tube muon chamber with iron absorbers. For hadrons, the dE/dx (TOF) resolution is $\sim 7.1\%$ (154 ps), providing $K/\pi$ separation of greater than $2\sigma$ (standard deviation) for particle momenta below 0.75 (1.07) GeV/c.

The data sample was collected from $e^+e^-$ collisions at a center-of-mass energy $\sqrt{s} \sim 10.6$ GeV. The total integrated luminosity of the sample is $1.58 fb^{-1}$, corresponding to $\sim 1.44 \times 10^6$ $\tau^+\tau^-$ events. The candidate events are required to contain two or four charged tracks with zero net charge. To reject beam-gas events, the distance of closest approach of each track to the IP must be within 5 mm transverse to the beam and 5 cm along the beam direction. Two-photon and Bhabha events are suppressed by a requirement on the total visible energy: $0.25 < E_{vis}/\sqrt{s} < 0.85$. The two-photon background is further suppressed by demanding that the measured transverse momentum of the event be greater than 1.0 GeV/c. To ensure reliable particle identification, the momentum vector of the kaon candidate, $\vec{p}_K$, must have magnitude less than 0.7 GeV/c and point into the barrel region, $|\cos\theta| < 0.81$, where $\theta$ is the polar angle with respect to the beam. The total momentum vector of the charged particle(s) recoiling against the kaon candidate, $\vec{p}_{tag}$, is used to divide the event into two hemispheres, forming a "1+1" or "1+3" *charged* track topology. The opening angle between $\vec{p}_K$ and $\vec{p}_{tag}$ must be greater than $120^0$. For the "1+1" topology, the requirement $p_{tag} > 1.0$ GeV/c is imposed in order to suppress the two-photon background. The hadronic background is suppressed by a requirement on the total invariant mass of the charged particles and photons in each hemisphere: $M < 1.7$ GeV/c$^2$. Also, the opening angle $\alpha$ between $\vec{p}_K$ ($\vec{p}_{tag}$) and the charged particles and photons in the same hemisphere must satisfy $\cos\alpha > 0.3$ (0.7).

Kaon candidates are identified using dE/dx and TOF information. The dE/dx measurement must be within $2\sigma$ of that expected for a kaon and, if good TOF information is available,[5] the TOF must be within $2\sigma$ of that expected for a kaon and greater than $4\sigma$ away from that for a pion. If $p_K > 0.4$ GeV/c but good TOF information is not available, the event is eliminated. No kaon candidate is allowed to be identified as an electron using the calorimeter information. To further suppress the hadronic background, any event of "1+3" topology is also discarded if both hemispheres contain a kaon candidate. A sample of 230 events satisfies these selection criteria.

Candidates for exclusive decay modes with $\pi^0$'s are identified using the calorimeter information in the kaon hemisphere. A photon candidate is defined as a crystal cluster with a minimum energy of 60 MeV in the angular region $|\cos\theta| < 0.71$ or 100 MeV in the region $0.71 < |\cos\theta| < 0.95$. This cluster must be isolated by at least 30 cm from the projection of any charged track unless its energy is greater than 300 MeV. To further discriminate against fake photons, a sub-sample of "high quality" photons is defined as those passing the isolation cut *and* having either an energy which is above 300 MeV or a lateral profile of energy deposition consistent with that expected of a photon. Only "high quality" photon candidates are included in the selection criteria described earlier. For the decay $\tau^- \to K^-\nu_\tau$, no "high quality" photons are allowed in the kaon hemisphere and for $\tau^- \to K^-\pi^0\nu_\tau$ ($K^-2\pi^0\nu_\tau$), there must be two (four) photons. In order to minimize the dependence on the modeling of showers, $\pi^0$ candidates accompanied by photons not classified as "high quality" are also accepted.

The distribution of the photon-pair invariant mass, $M_{\gamma\gamma}$, is shown in Fig. 1(a) for the $\tau^- \to K^-\pi^0\nu_\tau$ candidates. The $\pi^0$ signal corresponds to the first direct observation of this decay mode. The $K\gamma\gamma$ invariant mass spectrum of events with $\Delta M = |M_{\gamma\gamma} - M_{\pi^0}| <$ 15 MeV/c$^2$ ($\sim 3\sigma$) exhibits the resonant shape of $K^*(892)$, as shown in Fig. 1(b). There is no indication of non-resonant production as the eight events with $1.0 < M_{K\gamma\gamma} < 1.7$ GeV/c$^2$ are consistent with the expectation of $6.1 \pm 0.4$ events. The number of $K^-\pi^0$ events is extracted by fitting the $M_{\gamma\gamma}$ spectrum using a Gaussian with a long low-mass tail over a flat background. The width of the $\pi^0$ signal is constrained to the Monte Carlo expectation (see below). For the decay $\tau^- \to K^-2\pi^0\nu_\tau$, three "1+1" events contain two exclusive $\pi^0$ candidates ($\Delta M < 10$ MeV/c$^2$); the combinatoric background is estimated to be $0.25 \pm 0.56$ events using the two-dimensional $M_{\gamma\gamma}$ side-bands. Events of "1+3" topology are ignored due to the much higher hadronic contamination. The number of events for the various decays are summarized in Table 1.

The detection efficiency is calculated using a Monte Carlo simulation. The KORALB program is used to generate  pairs according to the standard electroweak theory, including $\alpha^3$ radiative corrections.[6] The detector response is simulated using the GEANT program.[7] Detector activity not attributable to the $e^+e^-$ interaction is modeled by embedding random trigger events into the generated events. The Monte Carlo kaon identification efficiency has been calibrated as a function of momentum using $D^{*+} \to D^0\pi^+ \to K^-\pi^+\pi^+$ decays from the data. The simulation reproduces the data quite well. For example, comparisons of the observed momentum spectra for three decay modes with the simulation are shown in Fig. 2. The Monte Carlo calculation yields the detection efficiencies[8] and backgrounds shown[9] in Tables 1 and 2.

The misidentification probability is calculated empirically as a function of momentum and polar angle using isolated pions from $K_S$ decays in hadronic events. The same probability is also used for the small $e$ and $\mu$ backgrounds, except the $e$ misidentification probability for dE/dx which is extracted from radiative Bhabhas and photon conversion. Two-photon backgrounds are determined using Monte Carlo simulations.[10] The hadronic background in the "1+1" topology is calculated using the Lund Monte Carlo program.[11] The background in the "1+3" topology is estimated by using this program to predict the *shape* of the total invariant mass spectrum of the tag hemisphere, but without the kaon identification cuts in order to increase statistics. The spectrum is then normalized to that observed in the data in the non-$\tau$ region, $M_{tag} > 1.8$ GeV/c$^2$. This estimate is consistent with the absolute prediction from the Lund program, which has a larger error.

After correcting for the background and detection efficiency, the branching ratios are extracted by normalizing to the luminosity and cross section. The world average topological branching ratios[3] are used in extracting the results, which are summarized in Table 1.

There are several sources of systematic errors as shown in Table 3. The uncertainty in the kaon identification efficiency is due to the limited sample of $D^*$ events. The uncertainty in the misidentification background from pions includes the statistical error due to the limited sample of misidentified pions from $K_S$ and the dependence of the misidentification probability on the isolation criteria. The reliability of the misidentification calculation has been verified by measuring the branching ratios using kaons in the momentum region 0.7-1.0 GeV/c which has $\sim 50\%$ misidentification background. The systematic error in the detection efficiency due to the uncertainty in the simulation of photons and hadronic in-teractions is estimated by varying the photon selection criteria.[12] The uncertainty in the acceptance due to the selection criteria is estimated by comparing the various kinematic distributions in the data with the expectations. The systematic error in the detection efficiency due to the uncertainty in the modeling of the Cabibbo-suppressed decays has been investigated by assuming different resonant substructures in the decays. The systematic error in the detection efficiency and migration background correction due to uncertainties in the $\tau$ decay branching ratios is estimated by changing the branching ratios within their uncertainties (Ref. 3 and Table 2). The uncertainty in the trigger simulation is studied by comparing branching ratios determined with various requirements on the trigger logic. The systematic error in the hadronic background calculation is dominated by the statistical error due to the paucity of events with high invariant mass. The inclusive branching ratio has also been measured with an electron or muon tag, for which the background is negligible; these measurements are consistent with the results from the generic tags.

The branching ratios from the two topologies agree, indicating the reliability of the measurement since some of the systematic errors are different. Combining the two samples, with the independent systematic errors added in quadrature, yields the final results, $B(\tau^- \to K^- \geq 0 \text{ neutrals } \nu_\tau) = (1.60 \pm 0.12 \pm 0.19)\%$, $B(\tau^- \to K^-\nu_\tau) = (0.66 \pm 0.07 \pm 0.09)\%$, $B(\tau^- \to K^-\pi^0\nu_\tau) = (0.51 \pm 0.10 \pm 0.07)\%$, $B(\tau^- \to K^-2\pi^0\nu_\tau) = (0.14 \pm 0.10 \pm 0.03)\%$ or $< 0.3\%$ at the 90% confidence level,[13] where the first error is statistical and the second is systematic.

In conclusion, the branching ratios for the dominant Cabibbo-suppressed decays of the $\tau$ lepton have been measured. The results are consistent with the Standard Model expectations[2] and previous measurements[3], but significantly more precise. The decay $\tau^- \to K^-\pi^0\nu_\tau$ is observed for the first time with exclusive $\pi^0$ reconstruction and the $K\pi^0$ mass spectrum is consistent with saturation by the $K^*(892)$ resonance. The difference between the inclusive and the sum of the two exclusive branching ratios is consistent with the presence of the other decay modes assumed. A new limit on $\tau^- \to K^-2\pi^0\nu_\tau$ has been set.

We gratefully acknowledge the effort of the CESR staff in providing us with excellent luminosity and running conditions. This work was supported by the National Science Foundation, the U.S. Dept. of Energy, the Heisenberg Foundation, the SSC Fellowship program of TNRLC, and the A.P. Sloan Foundation.

## REFERENCES

1. In this paper, charge conjugate states are implied.
2. Y. S. Tsai, Phys. Rev. **D4**, 2821 (1971); H. B. Thacker and J. J. Sakurai, Phys. Lett. **36B**, 103 (1971).
3. K. Hikasa *et al.*, Phys. Rev. **D45**, Part 2 (1992).
4. Y. Kubota *et al.*, Nucl. Instr. Method **A320**, 66 (1992).
5. The requirements on TOF, such as minimum pulse height, reduce the detection efficiency by $\sim 20\%$.
6. S. Jadach and Z. Was, Comp. Phys. Comm. **36**, 191 (1985); *ibid.* **64**, 267 (1991); S. Jadach, J. H. Kuhn, and Z. Was, *ibid.* **64**, 275 (1991).
7. R. Brun *et al.*, CERN Report No. CERN-DD/EE/84-1, 1987 (unpublished).
8. The decay $\tau^- \to K^-K^0\nu_\tau$ is modeled using a phase space distribution with a V-A weak interaction. The decay $\tau^- \to K^-\eta\nu_\tau$ has been neglected (M. Artuso *et al.*, Phys. Rev. Lett. **69**, 3278 (1992).)
9. The decay $\tau^- \to K^-K^0\nu_\tau$ accounts for $\sim 90\%$ of the migration background in $\tau^- \to K^-\nu_\tau$. The migration background in $\tau^- \to K^-\pi^0\nu_\tau$ has three components: $\tau^- \to K^-2\pi^0\nu_\tau$ ($\sim 23\%$), $\tau^- \to K^-K^0\nu_\tau$ ($\sim 8\%$), and $\tau^- \to K^-K^0\pi^0\nu_\tau$ ($\sim 69\%$).
10. J. Vermasseren, Nucl. Phys. **B229**, 347 (1983).
11. T. Sjostrand and M. Bengtsson, Comp. Phys. Comm. **43**, 367 (1987).
12. M. Procario *et al.*, Phys. Rev. Lett. **70**, 1207 (1993).
13. The branching ratio for $\tau^- \to K^-2\pi^0\nu_\tau$ includes a contribution from $\tau^- \to K^-K^0\nu_\tau \to K^-2\pi^0\nu_\tau$.

Table 1. Summary of the signal, background, branching ratio, and detection efficiency. The errors are statistical only. All upper limits are at the 90% confidence level, based on the prediction of zero event.

| Decay | $K \geq 0$ neutrals | | $K$ | | $K\pi^0$ | | $K2\pi^0$ |
|---|---|---|---|---|---|---|---|
| Topology | 1+1 | 1+3 | 1+1 | 1+3 | 1+1 | 1+3 | 1+1 |
| Data | 168 | 62 | 89 | 37 | $35.3 \pm 6.7$ | $9.2 \pm 3.6$ | 3 |
| Misid. | $21.3 \pm 0.1$ | $6.3 \pm 0.1$ | $12.4 \pm 0.1$ | $4.3 \pm 0.1$ | $3.7 \pm 0.2$ | $0.7 \pm 0.1$ | $0.29 \pm 0.02$ |
| Combin. | - | - | - | - | - | - | $0.25 \pm 0.56$ |
| Migration | - | - | $8.0 \pm 1.0$ | $2.2 \pm 0.5$ | $2.0 \pm 0.2$ | $0.7 \pm 0.1$ | - |
| $e^+e^- \to q\bar{q}$ | $< 3.7$ | $1.4 \pm 1.9$ | $< 0.6$ | $0.1 \pm 0.2$ | $< 0.8$ | $0.1 \pm 0.2$ | $< 0.2$ |
| $\gamma\gamma \to ff$ | $0.5 \pm 0.2$ | $0.1 \pm 0.1$ | $0.4 \pm 0.2$ | $0.1 \pm 0.1$ | $0.1 \pm 0.1$ | $< 0.1$ | - |
| Eff (%) | $0.39 \pm 0.01$ | $0.72 \pm 0.03$ | $0.45 \pm 0.02$ | $0.95 \pm 0.06$ | $0.23 \pm 0.01$ | $0.39 \pm 0.03$ | $0.077 \pm 0.003$ |
| B (%) | $1.57 \pm 0.14$ | $1.70 \pm 0.24$ | $0.63 \pm 0.09$ | $0.72 \pm 0.14$ | $0.53 \pm 0.12$ | $0.44 \pm 0.21$ | $0.14 \pm 0.10$ |

Table 2. Summary of the branching ratios ($10^{-4}$) and detection efficiencies ($10^{-4}$) used as input in calculating the branching ratio for $\tau^- \to K^- \geq 0$ neutrals $\nu_\tau$.

| Decay | $K$ | $K\pi^0$ | $K2\pi^0$ | $KK^0$ | $KK^0\pi^0$ |
|---|---|---|---|---|---|
| B | $66 \pm 7$ | $51 \pm 10$ | $10 \pm 5$ | $20 \pm 10$ | $20 \pm 10$ |
| $\epsilon(1+1)$ | $50 \pm 2$ | $36 \pm 1$ | $29 \pm 1$ | $41 \pm 4$ | $11 \pm 1$ |
| $\epsilon(1+3)$ | $106 \pm 7$ | $58 \pm 2$ | $23 \pm 1$ | $73 \pm 13$ | $19 \pm 2$ |

Table 3. Summary of systematic errors (%).

| Decay | $K \geq 0$ neutrals | $K$ | $K\pi^0$ | $K2\pi^0$ |
|---|---|---|---|---|
| Identification | 6 | 6 | 6 | 6 |
| Misid | 2 | 2 | 2 | 2 |
| $\epsilon$(photon) | 4 | 4 | 8 | 16 |
| Acceptance | 4 | 4 | 4 | 4 |
| Decay Model | 1 | – | 1 | 6 |
| BR | 5 | 5 | 3 | – |
| Trigger (1+1) | 7 | 7 | 3 | 3 |
| $e^+e^- \to q\bar{q}$ | 1 | 1 | 1 | 2 |
| Luminosity | 1.5 | 1.5 | 1.5 | 1.5 |
| Cross-section | 1 | 1 | 1 | 1 |

# FIGURE CAPTIONS

1. (a) The invariant mass spectrum of the photon pair in $\tau^- \to K^-\pi^0\nu_\tau$ events. The curve shows a fit to the spectrum. (b) The invariant mass spectrum of $K^-\gamma\gamma$ combinations ($|M_{\gamma\gamma} - M_{\pi^0}| < 15$ MeV/c$^2$). The histogram shows the Monte Carlo expectation, including the background (dashed).

2. The momentum spectrum of the kaons in: (a) $\tau^- \to K^- \geq 0$ neutrals $\nu_\tau$, (b) $\tau^- \to K^-\nu_\tau$, (c) $\tau^- \to K^-\pi^0\nu_\tau$. The histogram shows the Monte Carlo expectation, including the background (dashed).